\documentclass[fp, twocolumn]{jpsj/jpsj3}

\usepackage{mathtools}
\usepackage{color}
\usepackage{subfigure}
\usepackage[T1]{fontenc}

\begin{document}

\title{Uniformity Bias in Ground-State Sampling Induced by Replica Alignment in Quantum Monte Carlo for Quantum Annealing}

\author{%
Naoki Maruyama$^{1,4}$, Masayuki Ohzeki$^{1,2,3,4}$ and Kazuyuki Tanaka$^{1}$
}
\inst{%
$^1$Graduate School of Information Science, Tohoku University, Sendai 980-8579, Japan,
$^2$Department of Physics, Institute of Science Tokyo, Tokyo 152-8551, Japan,
$^3$Semiconductors and Informatics, Kumamoto University, Kumamoto 860-8555, Japan,
$^4$Sigma-i Co., Ltd., Tokyo 108-0075, Japan
}

\abst{%
Quantum annealing (QA) with a transverse field often fails to sample degenerate ground states fairly, limiting applicability to problems requiring diverse optimal solutions.
Although Quantum Monte Carlo (QMC) is widely used to simulate QA, its ability to reproduce such unfair ground-state sampling remains unclear because stochastic and coherent quantum dynamics differ fundamentally.
We quantitatively evaluate how accurately QMC reproduces the sampling bias in QA by comparing the final ground-state distributions from the QMC master equation and the Schrödinger equation.
We find QMC tends to produce uniform ground-state probabilities, unlike QA’s biased distribution, and that this uniformity bias strengthens as annealing proceeds.
Our analysis reveals that this bias originates from replica alignment—the dominance of configurations in which all Trotter replicas coincide—caused by the energetic suppression and entropic reduction of kink configurations (replica mismatches).
These findings clarify a fundamental limitation of discrete-time QMC in faithfully simulating QA dynamics, highlighting the importance of replica correlations and transition rules in achieving realistic ground-state sampling.
}

\maketitle

% Introduction
\section{Introduction}

% Quantum annealing
Quantum annealing (QA) utilizes quantum fluctuations to solve combinatorial optimization problems~\cite{kadowaki1998,farhi2001}.
Commercial hardware implementations such as those by D-Wave Systems have steadily advanced in the number of qubits and control accuracy~\cite{johnson2010,dattani2019,boothby2020}, and QA has been applied to optimization problems across various fields~\cite{ide2020,neukart2017,shikanai2025,ohzeki2018a,haba2022}.

% Unfair sampling of degenerate ground states
In many applications that exploit multiple optimal solutions, including SAT filters~\cite{weaver2012,azinovic2017} and machine learning models~\cite{hinton2002,eslami2014}, it is important to obtain these solutions uniformly.
However, QA with a transverse field is not guaranteed to sample degenerate ground states fairly, and this behavior, often called unfair sampling, has been reported both theoretically~\cite{matsuda2009,konz2019} and experimentally on D-Wave devices~\cite{mandra2017}.
Understanding the mechanism of this unfair sampling is crucial for extending the practical applicability of QA beyond simple optimization tasks.

% How to simulate QA
To investigate such properties, various simulation techniques have been developed.
While the Schrödinger equation provides the most faithful description of QA dynamics, its numerical solution is limited to small systems.
Quantum Monte Carlo (QMC) methods are widely used to study QA on larger systems~\cite{santoro2002,isakov2016,jiang2017a}.
QMC maps the quantum system to a classical one through the Suzuki–Trotter decomposition and samples equilibrium configurations of the resulting effective Hamiltonian.
Nevertheless, QMC and QA follow fundamentally different dynamics—stochastic Markovian evolution versus coherent quantum evolution—and this difference can lead to discrepancies in behavior.
Previous studies have pointed out that QMC does not always reproduce QA accurately.
For example, QMC can fail to equilibrate within polynomial time~\cite{hastings2013}, and incoherent quantum tunneling can outperform classical QMC escape processes~\cite{andriyash2017}.
While some comparisons between QMC and D-Wave annealers have shown qualitative similarities~\cite{boixo2014,denchev2016}, other works have reported clear deviations~\cite{albash2015,albash2015b}.
QMC has also been used for systems with degeneracy~\cite{bonca1993} and in studies addressing unfair sampling in QA~\cite{matsuda2009,konz2019}.

% Arguments
Despite these efforts, it remains unclear which aspects of QA QMC can faithfully reproduce.
In particular, the effect of the Trotter number, which controls the discretization error of the Suzuki–Trotter decomposition, on the fairness of ground-state sampling has not been fully clarified.
A central question is how the discretization and stochastic updates in QMC affect the biased ground-state sampling characteristic of QA.

This study focuses on ground-state sampling and quantitatively examines how accurately QMC can reproduce the unfairness observed in QA.
By comparing the results of the discrete-time QMC master equation with those obtained from the Schrödinger equation, we find that the simulation accuracy of QMC deteriorates because the probabilities of degenerate ground states tend to become uniform as the annealing process proceeds.
This phenomenon is referred to as the uniformity bias in QMC.

We identify the origin of this bias as the alignment of replicas along the Trotter axis.
As the coupling between replicas strengthens during annealing, configurations in which all replicas share the same spin pattern become dominant.
This alignment suppresses configurations that contain mismatches between adjacent replicas, referred to as kinks in this work, which are essential for reproducing the sampling bias intrinsic to QA.
Two mechanisms contribute to this alignment: an energetic suppression of kink configurations caused by ferromagnetic coupling between replicas, and an entropic reduction due to the limited number of admissible kink positions at smaller Trotter numbers.
We also find that the choice of transition rule in QMC, such as the Metropolis or heat-bath update, affects the rate of convergence to the stationary distribution and leads to differences in the degree of uniformity bias.

These results reveal a fundamental limitation of discrete-time QMC in reproducing QA’s sampling behavior.
They clarify that the suppression of kinks through replica alignment is the main factor that drives the discrepancy between QMC and QA.
Our findings suggest that increasing the Trotter number toward the continuous-time limit can reduce this bias and that alternative transition rules may help design QMC schemes that either better emulate QA or intentionally promote uniform sampling when desired.

% Methods
\section{Methods}

% Quantum annealing
QA is an algorithm that solves combinatorial optimization problems by harnessing quantum fluctuations.
A combinatorial optimization problem is equivalent to an Ising model, whose cost function is represented by the following Hamiltonian.
\begin{equation}
\label{eq:ising_H0}
H_0 = -\sum_{i=1}^{N} \sum_{j=1}^{N} J_{ij} \, \sigma_i \, \sigma_{j},
\end{equation}
where $\sigma_i \in \{1,-1\}$ is the $i$-th spin variable, $N$ is the number of spins, and $J_{ij}$ denotes the interaction strength between $i,j$-th spins.
We consider transverse-field QA with the time-dependent Hamiltonian
\begin{equation}
\label{eq:anneal_Ht}
\hat{H}(t)=\frac{t}{\tau} \, \hat{H}_0 \left( \{ \hat{\sigma}_i^{z} \} \right) - \left(1-\frac{t}{\tau}\right) \sum_{i=1}^N \hat{\sigma}_i^x,
\end{equation}
where $\hat{\sigma}_i^{x},\hat{\sigma}_i^{z}$ are the $x,z$ components of the Pauli matrices for the $i$-th spin, and $\tau$ is the annealing time.
In the computation process of QA, an equal superposition over all computational basis states is prepared at the initial time $t=0$, and the quantum effect is gradually reduced until the final time $t=\tau$.
If the time evolution is sufficiently slow, the adiabatic theorem guarantees that the ground state is obtained at the end \cite{morita2008}.
The standard QA enables the system to evolve according to the Schrödinger equation.
\begin{equation}
\label{eq:schrodinger}
i \, \frac{d \, \Psi(t)}{dt} = \hat{H}(t) \, \Psi(t),
\end{equation}
where $\Psi(t)$ is the state vector at time $t$.
Our target in the QA simulation is the final state obtained by time evolution under the Schrödinger equation.
In other words, the closer the terminal distribution QMC produces to this state, the higher the simulation accuracy.
Note that our goal does not align with the time evolution.
However, we later examine the probability dynamics to obtain ground states, which will help us better understand the final states.

% Quantum Monte Carlo method
QMC simulates quantum systems using Monte Carlo methods and can be applied to simulate QA.
The partition function of the transverse-field Ising model can be written as $Z = \operatorname{Tr} \left\{\exp (-\beta \, \hat{H} (t)) \right\}$, where $\beta$ is the inverse temperature.
Applying the Suzuki-Trotter decomposition to this partition function yields a classical Ising model in one higher dimension consisting of replicas of the original problem, with an effective Hamiltonian.
\begin{equation}
\label{eq:heff}
H_{\mathrm{eff}}(\sigma)=\frac{s}{M} \sum_{k=1}^M H_0\left(\sigma_k\right) - J^{\star} \sum_{k=1}^M \sum_{i=1}^N \sigma_{i,k} \, \sigma_{i,k+1},
\end{equation}
and the partition function $Z \simeq \operatorname{Tr} \left\{\exp (-\beta \, H_{\mathrm{eff}}(\sigma)) \right\}$.
Here, $\sigma_{i,k} \in \{1,-1\}$ is the $i$-th spin variable in the $k$-th replica, $M$ is the Trotter number, and $J^{\star}=\frac{1}{2 \beta} \ln \coth \left(\frac{\beta}{M} \left( 1 - s \right) \right)$, $s=t/\tau$.
We set $\beta=M$ so that $\beta / M$ remains constant and the dimensionless Trotter coupling $\beta J_{\star}$ depends only on $s$.
The larger $M$ reduces the decomposition error, bringing the mapping closer to the original quantum system.
We impose periodic boundary conditions along the Trotter axis.
Initially, the inter-replica coupling is weak, and each replica explores independently.
As the coupling is strengthened, the spin configurations across replicas align.

In QMC, the system evolves according to the following master equation:
\begin{equation}
\label{eq:master_cont}
\frac{dP(t)}{dt} = \mathcal{R}(t) P(t),
\end{equation}
where $P(t)$ is the state-probability vector and is satisfied with $\sum_{\sigma} P(\sigma, t)=1$.
$\mathcal{R}_{\sigma\sigma^{\prime}} (t)$ represents the transition rate from state $\sigma^{\prime}$ to $\sigma$ per unit time.
While Eq. (\ref{eq:schrodinger}) and (\ref{eq:master_cont}) have a similar form, how closely their final states correspond to each other is nontrivial.

We use the discrete-time form instead of the continuous-time form above.
\begin{equation}
\label{eq:master_discrete}
\begin{aligned}
P(\sigma, t+\Delta t) &= \left(1-\sum_{\sigma^{\prime}(\neq \sigma)} w_{\sigma \to \sigma^{\prime}} \, \Delta t\right) \, P(\sigma, t) \\
&\quad+\sum_{\sigma^{\prime}(\neq \sigma)} w_{\sigma^{\prime} \to \sigma} \, \Delta t \, P\left(\sigma^{\prime},  t\right),
\end{aligned}
\end{equation}
where $w_{\sigma \to \sigma^{\prime}}$ is the transition probability per unit time from state $\sigma$ to $\sigma^{\prime}$.
The first term represents the probability of remaining in state $\sigma$, and the second term the probability of transitioning from state $\sigma^{\prime}$ to $\sigma$.
In Eq. (\ref{eq:master_discrete}), we adopt the convention that $w_{\sigma \to \sigma_{\prime}}$ denotes only the acceptance probability for single-spin-flip neighbors, hence $0 \leq w_{\sigma \to \sigma^{\prime}} \leq 1$.
Since each configuration has exactly $NM$ neighbors with the Hamming distance of one, we choose $\Delta t$ such that $\sum_{\sigma^{\prime}(\neq \sigma)} w_{\sigma \to \sigma^{\prime}}(t) \Delta t \leq 1$.
This reduces to $\Delta t \leq 1 /(N M)$ for single-spin flips.
We set $\Delta t = 0.05$, which satisfies this bound for all sizes used.
We also can write $P(t+\Delta t)=\mathcal{L}(t)\,P(t)$, where $\mathcal{L}_{\sigma\sigma}=1-\sum_{\sigma^{\prime}(\neq \sigma)} w_{\sigma \to \sigma^{\prime}}\,\Delta t$ and $\mathcal{L}_{\sigma\sigma^{\prime}}=w_{\sigma^{\prime} \to \sigma}\,\Delta t$ for $\sigma^{\prime}\neq\sigma$.
Starting from a suitable initial distribution $P(0)$, we compute $\mathcal{L}(t)$ at each time and multiply it by $P(t)$ to obtain the exact time evolution of probabilities in QMC.
We set the initial distribution to be uniform over all states.

% Transition rules satisfying the detailed balance condition
A transition probability in the master equation is typically chosen to satisfy the detailed balance condition, which guarantees convergence to the stationary distribution.
Representative choices satisfying the condition are the Metropolis method \cite{metropolis1953} and the heat-bath (Glauber) method \cite{glauber1963}, respectively given as
\begin{equation}
\label{eq:transition_rules_metropolis}
w_{\sigma \to \sigma^{\prime}} = \min \left(1, e^{-\beta\Delta E}\right)
\end{equation}
\begin{equation}
\label{eq:transition_rules_heat_bath}
w_{\sigma \to \sigma^{\prime}} = \left(1+e^{\beta\Delta E}\right)^{-1}
\end{equation}
where $\Delta E = H_{\mathrm{eff}}(\sigma^{\prime}) - H_{\mathrm{eff}}(\sigma)$ is the energy difference of the effective classical system.
The Metropolis method tends to have higher proposal acceptance than the heat-bath method \cite{janke2012}, although the difference is at most a factor of two.

% Results
\section{Results}

% Our toy model
This study examines the following toy model, similar to the one analyzed in another study \cite{chancellor2020}.
\begin{equation}
\label{eq:toy_H0}
H_0=-\sum_{i=1}^{N-1} \sigma_i \, \sigma_{i+1}-\sigma_1+\sigma_N
\end{equation}
This model has $N+1$ degenerate ground states: $|\uparrow\cdots\uparrow\uparrow\rangle, |\uparrow\cdots\uparrow\downarrow\rangle, \ldots, |\downarrow\cdots\downarrow\downarrow\rangle$, which are apart by the Hamming distance of one.
Under transverse-field QA, a bias arises that favors the center ground states with higher probability.
Furthermore, due to the symmetry of the system, one can obtain symmetric ground states such as $|\uparrow\uparrow\uparrow\rangle$ and $|\downarrow\downarrow\downarrow\rangle$ with equal probability.

% Dependence on Trotter number and annealing time
We define a metric to quantify simulation accuracy and the deviation from uniformity in ground-state sampling.
\begin{equation}
\label{eq:metric_D}
D \left(P_{\mathrm{QMC}}, P_{\mathrm{\ast}} \right) := \sum_{s=1}^{N_{\mathrm{GS}}} \left| P_{\mathrm{QMC}} \left(\sigma^s\right) - P_{\mathrm{\ast}} \left(\sigma^s\right) \right|,
\end{equation}
where $N_{\mathrm{GS}}$ is the number of degenerate ground states and $\sigma^s$ is the $s$-th ground state.
This metric is based on the total variation distance and sums the absolute differences between the probabilities in QMC and a comparator ($\ast$) across all ground states.
When we denote the distribution from the Schrödinger equation as $P_{\mathrm{SD}}$, the smaller the simulation error $D(P_{\mathrm{QMC}}, P_{\mathrm{SD}})$, the higher the accuracy of QMC in simulating QA.
We also use this metric to assess deviation from uniformity by comparing against the uniform distribution, $D(P_{\mathrm{QMC}}, P_{\mathrm{uniform}})$, where a larger value indicates less uniform sampling.

\begin{figure}[htb]
\centering
\subfigure[$D(P_{\mathrm{QMC}}, P_{\mathrm{SD}})$]{\includegraphics[width=\linewidth]{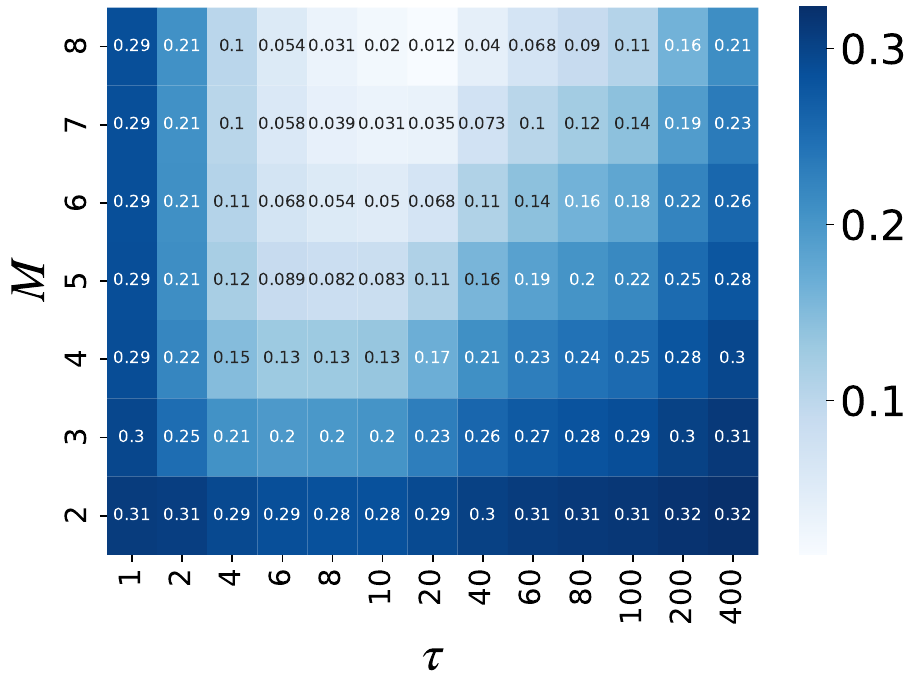}\label{fig:simulation_accuracy}}
\subfigure[$D(P_{\mathrm{QMC}}, P_{\mathrm{uniform}})$]{\includegraphics[width=\linewidth]{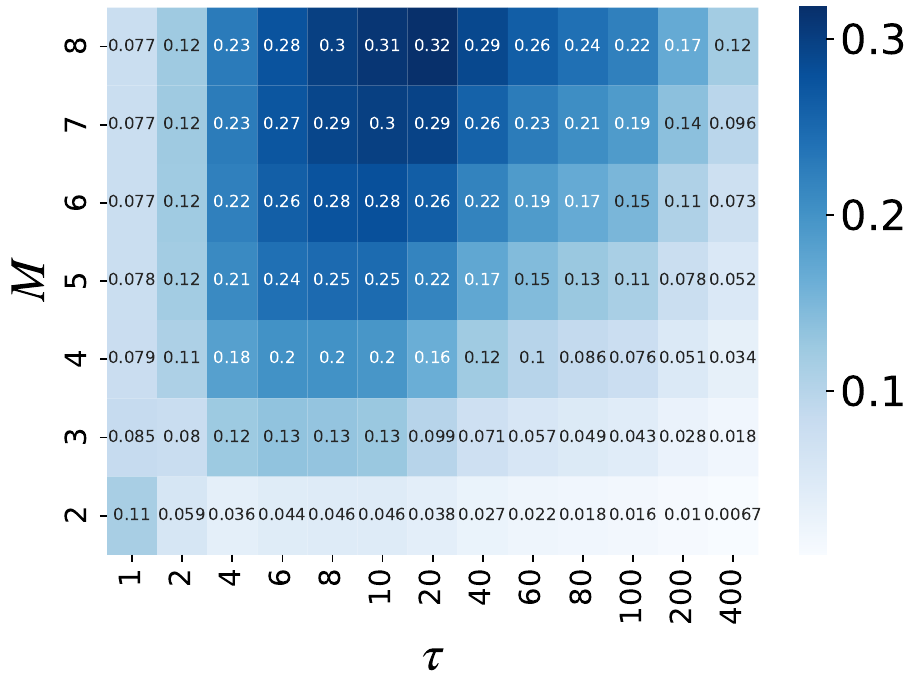}\label{fig:fairness}}
\caption{
(Color online)
(a) Simulation error $D(P_{\mathrm{QMC}}, P_{\mathrm{SD}})$ under Metropolis dynamics for each Trotter number and annealing time.
Larger values indicate lower simulation accuracy.
(b) Deviation from uniformity $D(P_{\mathrm{QMC}}, P_{\mathrm{uniform}})$ with the Metropolis method.
Smaller values indicate more uniform sampling over ground states.
}
\label{fig:results_overview}
\end{figure}

Figure \ref{fig:results_overview} shows how simulation accuracy depends on the Trotter number and the annealing time.
We set $N=2$, $M=2,3,\ldots,8$, and $\tau=1,2,\ldots,400$ and use the Metropolis method as the transition rule.
The reference distribution from the Schrödinger equation is taken at $\tau_{\text{SD}}=100$.
Hence, the heatmap should be interpreted as a score against a fixed reference rather than a strict like-for-like comparison at the same $\tau$.
Figure \ref{fig:simulation_accuracy} plots $D(P_{\mathrm{QMC}}, P_{\mathrm{SD}})$ for each $(\tau, M)$.
As expected from the Suzuki-Trotter decomposition, the accuracy improves overall as $M$ increases.
We observe a high-accuracy region centered around the upper center and low-accuracy regions in the left and lower-right areas.
The high-accuracy region is expanding toward the upper-right, indicating that longer annealing times require larger Trotter numbers to maintain accuracy.

To understand the three regions, we focus on the deviation from uniformity, i.e., the uniformity of QMC sampling.
As Figure \ref{fig:fairness} shows, $D(P_{\mathrm{QMC}}, P_{\mathrm{uniform}})$ decreases (i.e., sampling becomes more uniform) in the two regions where $D(P_{\mathrm{QMC}}, P_{\mathrm{SD}})$ increases (lower accuracy).
In addition, the simulation accuracy decreases in the left region (small annealing time) because the probability of obtaining any ground state deviates from 1.

\begin{figure}[htb]
\centering
\includegraphics[width=\linewidth]{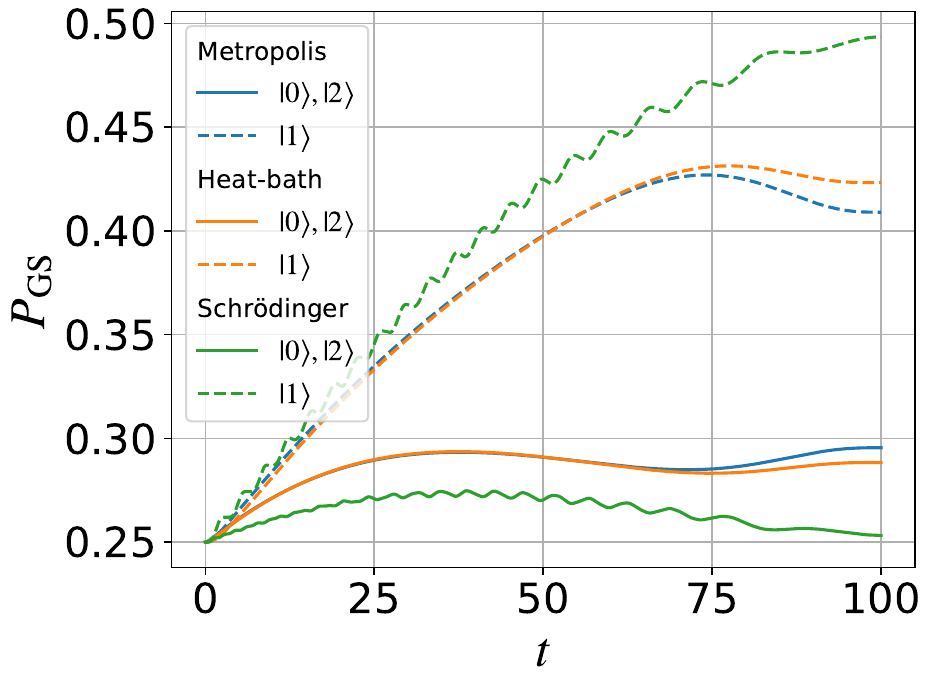}
\caption{
(Color online)
Time evolution of the probabilities of each ground state ($t$: MC steps; not physical time).
We present the results for the Schrödinger dynamics, Metropolis, and heat-bath methods.
Ground states are labeled in order, starting from $|\uparrow\cdots\uparrow\uparrow\rangle$.
}
\label{fig:time_vs_prob}
\end{figure}

We examine the system's time variation in QMC to understand why the simulation can become inaccurate in the above results.
Figure \ref{fig:time_vs_prob} shows the probability of each ground state for each step in QMC and the Schrödinger equation.
We set $N=2$, $M=6$, and $\tau=\tau_{\text{SD}}=100$ and adopt the Metropolis method as the transition rule.
Ground states are labeled in order, starting from $|\uparrow\cdots\uparrow\uparrow\rangle$.
While the trends are similar between QMC and Schrödinger dynamics up to around $\tau=70$, beyond that point, apparent differences emerge.
Specifically, under the Schrödinger equation, the gap between the probabilities of states $| 0 \rangle, | 2 \rangle$ and $| 1 \rangle$ continues to widen, whereas under QMC, the differences shrink toward the end.
We also include the heat-bath method result, showing slight differences near the final time.
We return to the gap in the transition rules later.

Examining the spin configuration of each replica reveals why QMC produces a more uniform result than Schrödinger dynamics.
We define the number of kinks as the total number of sites where the spin orientation differs between adjacent replicas.
\begin{equation}
\label{eq:kink}
K := \sum_{i,k} \frac{1-\sigma_{i, k} \sigma_{i, k+1}}{2}
\end{equation}
The pair between replicas $1$ and $M$ is included because we impose periodic boundary conditions along the Trotter axis.

Next, we identify that states containing kinks are energetically suppressed near the end of the annealing process.
For configurations with $K$ kinks, the unnormalized Boltzmann weight for couplings along the Trotter axis is given by $v(K) = \exp \left[\beta J^{\star} \sum_{i, k} \sigma_{i, k} \sigma_{i, k+1}\right]$.
Using $\sum_{i,k} \sigma_{i,k} \sigma_{i,k+1} = NM - 2K$, the relative weight with respect to the kink-free configuration is
\begin{equation}
\frac{v(K)}{v(0)} = \exp \left(-2 \beta J^{\star} K\right).
\end{equation}
This relative weight indicates that when its value is small, the state with kinks is suppressed, and a uniformity bias is strengthened.
It can be seen that the ferromagnetic interaction $J^{\star}$ between replicas contributes to the relative weight of kinks.
In this study, since $\beta/M=1$ is constant, both $J^{\star}$ and the relative weight are independent of $M$.
The dependence on $M$ observed in Figure \ref{fig:results_overview} becomes clear from a perspective other than energy, as discussed later.
We here denote $a = \frac{\beta}{M}(1-s)$, expressing $J^{\star}=\frac{1}{2 \beta} \ln \coth(a)$.
Near the end time, as $s \to 1 \ (a \to 0)$, the relative weight decreases by $\exp \left(-2 \beta J^{\star}\right) = \tanh a$ for each additional kink.
For small $a$, since $\tanh a \approx a$, the relative weight is suppressed by $a^K$, and the probability concentrates on the state with $K=0$.
Note that when $\beta/M$ is not constant, $a$ depends on $M$.
In $\beta$ fixed, the larger $M$ is, the greater the relative weight becomes, and we expect that states with kinks will be promoted.

\begin{figure}[htb]
\centering
\includegraphics[width=\linewidth]{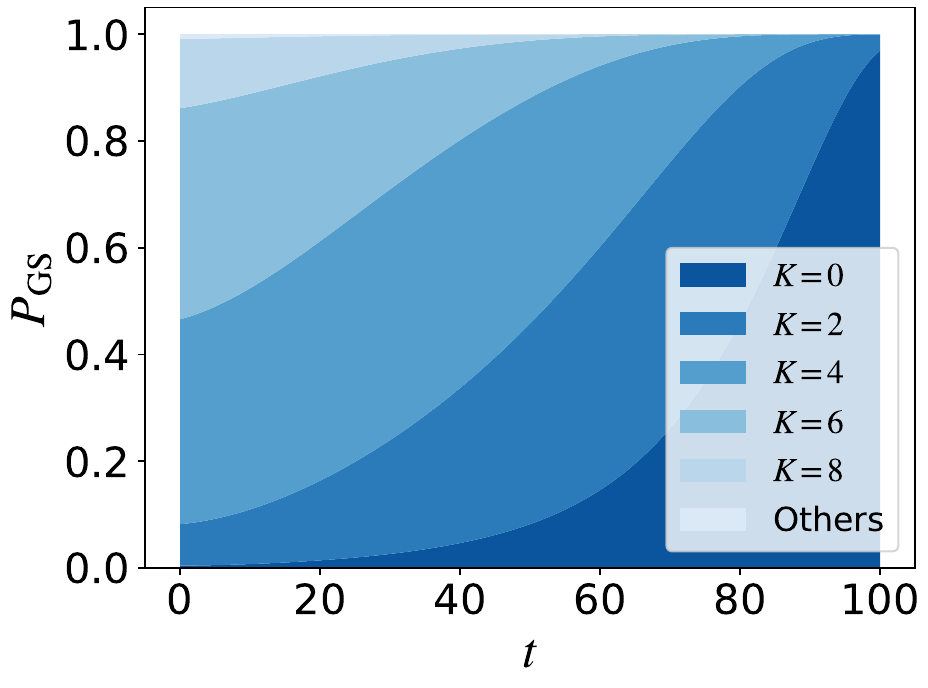}
\caption{
(Color online)
Time evolution of the probability of each kink-number sector in QMC under Metropolis dynamics.
}
\label{fig:kink}
\end{figure}

We show the time evolution of the number of kinks per $K$ in Figure \ref{fig:kink}, where $N=2$, $M=6$, and $\tau=100$, and the Metropolis method is used as the transition rule.
As time passes, the probability of obtaining a state with fewer kinks increases.
Primarily, the probability mass concentrates sharply on $K=0$ near the end.
This result is consistent with the analytical result concerning the relative weight of kinks mentioned above.

We also discuss the entropic contribution as the mechanism underlying the Trotter number dependence of the uniformity bias shown in Figure \ref{fig:results_overview}.
At $\beta/M=1$, the energy penalty per kink $\tanh (\frac{\beta}{M} (1-s))$ does not depend on $M$.
Therefore, the $M$-dependence of the kinks must stem from the number of possible locations for the kinks.
Denoting $K=\sum_{i, k} \mathbf{1}\left(\sigma_{i, k} \neq \sigma_{i, k+1}\right)$, and using the linearity of expectation, we have
\begin{equation}
\mathbb{E}\left[K\right]=N M q(s, M),
\end{equation}
where $q(s, M)=\mathbb{P} \left(\sigma_{i, k} \neq \sigma_{i, k+1}\right)$ is the probability of replica mismatch.

\begin{figure}[htb]
\centering
\includegraphics[width=\linewidth]{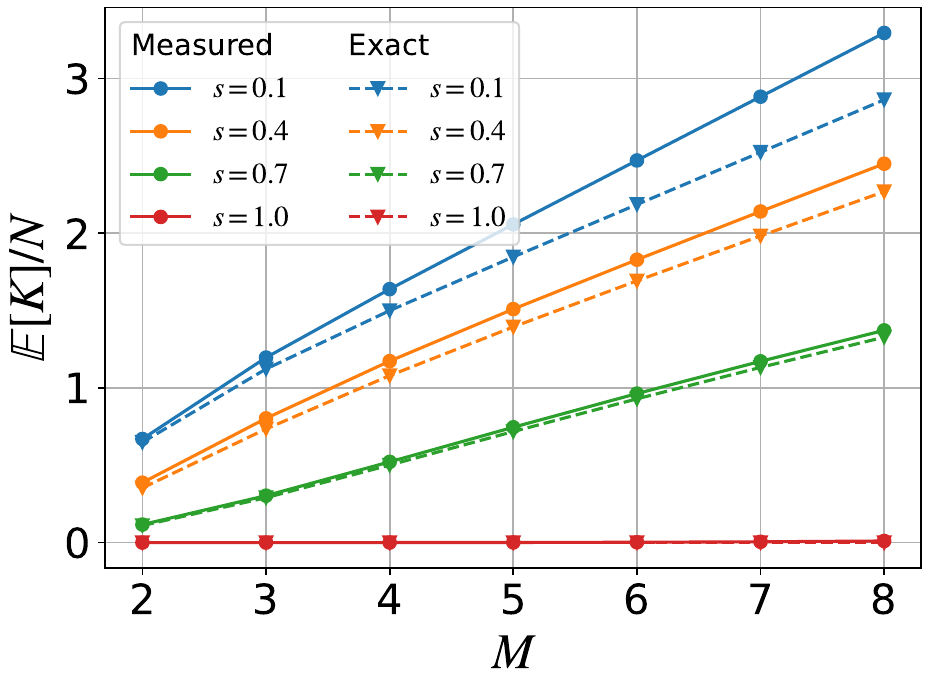}
\caption{
(Color online)
The Trotter number dependence of the expected number of kinks per one site at each time $s=0.1, 0.4, 0.7, 1.0$.
The measured values obtained from the QMC's master equation and the equilibrium values exactly derived from the partition function are plotted respectively.
}
\label{fig:expected_num_kinks}
\end{figure}

Figure \ref{fig:expected_num_kinks} shows that for each $s$, both the measurement results from the QMC's master equation and the equilibrium state rigorously computed from the partition function increase almost linearly with $M$.
This suggests that the primary $M$-dependence stems from the combinatorial factor $M$ ("entropic effect").
The exact curve always lies below the measured values.
We assume this because the finite annealing time causes a freeze-out, leaving slightly more kinks than in the equilibrium state.
Furthermore, the slope of each line decreases as $s$ increases.
The energy penalty for each kink $\tanh (\frac{\beta}{M} (1-s))$ increases, causing $q(s, M)$ to decrease.
Therefore, as $M$ decreases, fewer sites are available for kinks, resulting in fewer kinks and more substantial replica alignment, which is consistent with previous results.

\begin{figure}[htb]
\centering
\includegraphics[width=\linewidth]{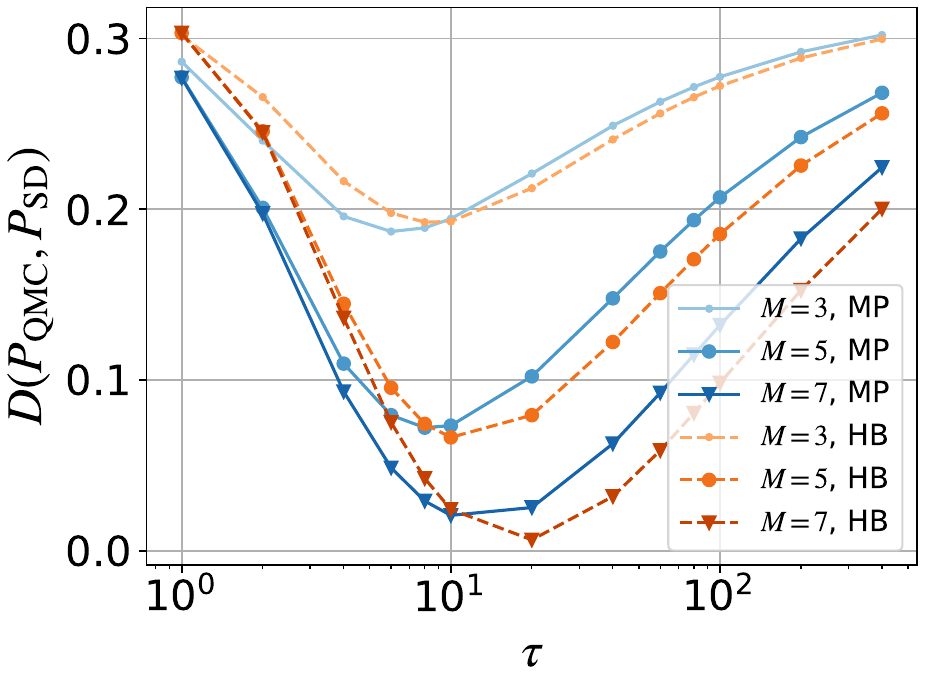}
\caption{
(Color online)
Dependence of simulation error $D(P_{\mathrm{QMC}}, P_{\mathrm{SD}})$ on annealing time and Trotter number for different transition rules.
The heat-bath method achieves a minimum error at larger $(M,\tau)$ values than the Metropolis method.
}
\label{fig:transitions_comparison}
\end{figure}

% Comparison of different transition rules
The choice of transition rule affects simulation accuracy by QMC as seen in Figure \ref{fig:time_vs_prob}.
Figure \ref{fig:transitions_comparison} shows the simulation error $D(P_{\mathrm{QMC}}, P_{\mathrm{SD}})$ across different Trotter numbers and annealing times for the Metropolis (MP) and heat-bath (HB) methods.
We set $N=2$, $M=3,5,7$, $\tau=1,2,\ldots,400$, and $\tau_{\text{SD}}=100$ (same as in Figure \ref{fig:results_overview}).
The conditions under which the error is minimized differ depending on the rules.
Specifically, the heat-bath method achieves its minimum error at larger $M$ and $\tau$ than the Metropolis method.
This difference is consistent with Metropolis updates' typically higher acceptance rate and the resulting difference in autocorrelation times \cite{pollet2004}.
Therefore, we infer that the rapid convergence to the aforementioned undesirable state causes the ground-state probabilities to approach a uniform distribution.

% Discussion
\section{Discussion}

% Summary of this study
We investigated how accurately QMC can simulate the fairness of degenerate ground-state sampling in QA.
By solving the QMC master equation and comparing with ground-state distributions by the Schrödinger equation, we found that QMC samples degenerate ground states more fairly than QA.
We demonstrated that the uniformity results from suppressing kinks between adjacent replicas.
Moreover, we identified energetic suppression and entropic effect as mechanisms driving the replica alignment.
Our analytical and numerical results show that the ferromagnetic couplings along the Trotter axis suppress the relative weight of kink configurations.
We further argued that smaller Trotter numbers reduce the number of available kink positions, making it easier for replicas to align.
Although we conducted experiments for specific models and sizes this time, it would be worthwhile to investigate these dependencies for a broader class of problems in the future, since the mechanism of our results is essentially general.

% Continuous time QMC
Based on our results, we expect that continuous-time QMC ($M\to\infty$) can improve the simulation accuracy \cite{rieger1998,heim2015}.
As shown in the above section, the alignment bias induced by discretization scales as a power of $\beta/M$ and therefore vanishes in the continuous-time limit.
Furthermore, as $M \rightarrow \infty$, the absence of constraints on where kinks can form is expected to resolve the replica alignment.
For example, the continuous-time QMC proposed in \cite{barash2024} is free from Trotter errors, and thus should suppress the bias toward the state where all replicas are aligned.

% Non-detailed balance MCMC
The observed differences in accuracy across transition rules suggest that there is room to reduce the uniformity bias or to intentionally promote fair sampling by designing the rule.
One might design the transition rule to avoid "over-optimizing" for the inefficient Trotter number.
On the other hand, classical algorithms specialized for fair sampling can be constructed by deliberately over-optimizing the transitions.
In addition, recent methods intentionally break detailed balance to accelerate Monte Carlo simulations \cite{manousiouthakis1999,ichiki2013}, whereas we employed standard rules here that preserve this condition.
We expect that such approaches can help improve the simulation accuracy for QA and the fairness of sampling.

% Acknowledgment
\section*{Acknowledgment}
This study was financially supported by programs for bridging the gap between R\&D and IDeal society (Society 5.0) and Generating Economic and social value (BRIDGE) and Cross-ministerial Strategic Innovation Promotion Program (SIP) from the Cabinet Office 23836436.

% Author contributions
\section*{Author contributions}
N.M. conceived of the presented idea and performed the experiments.
M.O. supervised the findings of this work.
K.T. helped to supervise the project.
All authors discussed the results and contributed to preparing the final manuscript.

* naoki.maruyama.p7@dc.tohoku.ac.jp
\bibliographystyle{jpsj/jpsj}
\bibliography{main}

\begin{thebibliography}{10}

\bibitem{kadowaki1998}
T.~Kadowaki and H.~Nishimori: Physical Review E {\bfseries 58} (1998) 5355.

\bibitem{farhi2001}
E.~Farhi, J.~Goldstone, S.~Gutmann, J.~Lapan, A.~Lundgren, and D.~Preda: Science {\bfseries 292} (2001) 472.

\bibitem{johnson2010}
M.~W. Johnson, P.~Bunyk, F.~Maibaum, E.~Tolkacheva, A.~J. Berkley, E.~M. Chapple, R.~Harris, J.~Johansson, T.~Lanting, I.~Perminov, E.~Ladizinsky, T.~Oh, and G.~Rose: Superconductor Science and Technology {\bfseries 23} (2010) 065004.

\bibitem{dattani2019}
N.~Dattani, S.~Szalay, and N.~Chancellor: arXiv:1901.07636 [quant-ph]  (2019).

\bibitem{boothby2020}
K.~Boothby, P.~Bunyk, J.~Raymond, and A.~Roy: arXiv:2003.00133 [quant-ph]  (2020).

\bibitem{ide2020}
N.~Ide, T.~Asayama, H.~Ueno, and M.~Ohzeki: 2020 {{International Symposium}} on {{Information Theory}} and {{Its Applications}} ({{ISITA}}), October 2020, pp. 91--95.

\bibitem{neukart2017}
F.~Neukart, G.~Compostella, C.~Seidel, D.~{von Dollen}, S.~Yarkoni, and B.~Parney: Frontiers in ICT {\bfseries 4} (2017).

\bibitem{shikanai2025}
R.~Shikanai, M.~Ohzeki, and K.~Tanaka: Journal of the Physical Society of Japan {\bfseries 94} (2025) 024001.

\bibitem{ohzeki2018a}
M.~Ohzeki, A.~Miki, M.~J. Miyama, and M.~Terabe.
\newblock Control of Automated Guided Vehicles without Collision by Quantum Annealer and Digital Devices.
\newblock https://arxiv.org/abs/1812.01532v2, December 2018.

\bibitem{haba2022}
R.~Haba, M.~Ohzeki, and K.~Tanaka: Scientific Reports {\bfseries 12} (2022) 17753.

\bibitem{weaver2012}
S.~A. Weaver, K.~J. Ray, V.~W. Marek, A.~J. Mayer, and A.~K. Walker: Journal on Satisfiability, Boolean Modeling and Computation {\bfseries 8} (2012) 129.

\bibitem{azinovic2017}
M.~Azinovi{\'c}, D.~Herr, B.~Heim, E.~Brown, and M.~Troyer: SciPost Physics {\bfseries 2} (2017) 013.

\bibitem{hinton2002}
G.~E. Hinton: Neural Computation {\bfseries 14} (2002) 1771.

\bibitem{eslami2014}
S.~M.~A. Eslami, N.~Heess, C.~K.~I. Williams, and J.~Winn: International Journal of Computer Vision {\bfseries 107} (2014) 155.

\bibitem{matsuda2009}
Y.~Matsuda, H.~Nishimori, and H.~G. Katzgraber: New Journal of Physics {\bfseries 11} (2009) 073021.

\bibitem{konz2019}
M.~S. K{\"o}nz, G.~Mazzola, A.~J. Ochoa, H.~G. Katzgraber, and M.~Troyer: Physical Review A {\bfseries 100} (2019) 030303.

\bibitem{mandra2017}
S.~Mandr{\`a}, Z.~Zhu, and H.~G. Katzgraber: Physical Review Letters {\bfseries 118} (2017) 070502.

\bibitem{santoro2002}
G.~E. Santoro, R.~Marto{\v n}{\'a}k, E.~Tosatti, and R.~Car: Science {\bfseries 295} (2002) 2427.

\bibitem{isakov2016}
S.~V. Isakov, G.~Mazzola, V.~N. Smelyanskiy, Z.~Jiang, S.~Boixo, H.~Neven, and M.~Troyer: Physical Review Letters {\bfseries 117} (2016) 180402.

\bibitem{jiang2017a}
Z.~Jiang, V.~N. Smelyanskiy, S.~V. Isakov, S.~Boixo, G.~Mazzola, M.~Troyer, and H.~Neven: Physical Review A {\bfseries 95} (2017) 012322.

\bibitem{hastings2013}
M.~B. Hastings and M.~H. Freedman: arXiv:1302.5733 [quant-ph]  (2013).

\bibitem{andriyash2017}
E.~Andriyash and M.~H. Amin: arXiv:1703.09277 [quant-ph]  (2017).

\bibitem{boixo2014}
S.~Boixo, T.~F. R{\o}nnow, S.~V. Isakov, Z.~Wang, D.~Wecker, D.~A. Lidar, J.~M. Martinis, and M.~Troyer: Nature Physics {\bfseries 10} (2014) 218.

\bibitem{denchev2016}
V.~S. Denchev, S.~Boixo, S.~V. Isakov, N.~Ding, R.~Babbush, V.~Smelyanskiy, J.~Martinis, and H.~Neven: Physical Review X {\bfseries 6} (2016) 031015.

\bibitem{albash2015}
T.~Albash, T.~R{\o}nnow, M.~Troyer, and D.~Lidar: The European Physical Journal Special Topics {\bfseries 224} (2015) 111.

\bibitem{albash2015b}
T.~Albash, I.~Hen, F.~M. Spedalieri, and D.~A. Lidar: Physical Review A {\bfseries 92} (2015) 062328.

\bibitem{bonca1993}
J.~Bon{\v c}a and J.~E. Gubernatis: Physical Review B {\bfseries 47} (1993) 13137.

\bibitem{morita2008}
S.~Morita and H.~Nishimori: Journal of Mathematical Physics {\bfseries 49} (2008) 125210.

\bibitem{metropolis1953}
N.~Metropolis, A.~W. Rosenbluth, M.~N. Rosenbluth, A.~H. Teller, and E.~Teller: The Journal of Chemical Physics {\bfseries 21} (1953) 1087.

\bibitem{glauber1963}
R.~J. Glauber: Journal of Mathematical Physics {\bfseries 4} (1963) 294.

\bibitem{janke2012}
W.~Janke: {\em Monte {{Carlo Simulations}} in {{Statistical Physics}} --- {{From Basic Principles}} to {{Advanced Applications}}} (WORLD SCIENTIFIC, December 2012), pp. 93--166.

\bibitem{chancellor2020}
N.~Chancellor, P.~J.~D. Crowley, T.~{\DJ}uri{\'c}, W.~Vinci, M.~H. Amin, A.~G. Green, P.~A. Warburton, and G.~Aeppli: arXiv:2006.07685 [quant-ph]  (2020).

\bibitem{pollet2004}
L.~Pollet, S.~M.~A. Rombouts, K.~Van~Houcke, and K.~Heyde: Physical Review E {\bfseries 70} (2004) 056705.

\bibitem{rieger1998}
H.~Rieger and N.~Kawashima.
\newblock Application of a Continous Time Cluster Algorithm to the {{Two-dimensional Random Quantum Ising Ferromagnet}}, February 1998.

\bibitem{heim2015}
B.~Heim, T.~F. R{\o}nnow, S.~V. Isakov, and M.~Troyer: Science {\bfseries 348} (2015) 215.

\bibitem{barash2024}
L.~Barash, A.~Babakhani, and I.~Hen: Physical Review Research {\bfseries 6} (2024) 013281.

\bibitem{manousiouthakis1999}
V.~I. Manousiouthakis and M.~W. Deem: The Journal of Chemical Physics {\bfseries 110} (1999) 2753.

\bibitem{ichiki2013}
A.~Ichiki and M.~Ohzeki: Physical Review E {\bfseries 88} (2013) 020101.

\end{thebibliography}

\end{document}